\documentclass[showpacs,twocolumn]{revtex4}
\usepackage[dvips]{epsfig}

\newcommand{\be}{\begin{equation}}
\newcommand{\ee}{\end{equation}}
\newcommand{\clF}{{\cal F}}
\newcommand{\clT}{{\cal T}}

\newcommand{\clW}{{\cal W}}
\newcommand{\tr}{{\rm Tr}\,}
\newcommand{\bea}{\begin{eqnarray}}
\newcommand{\eea}{\end{eqnarray}}
\newcommand{\hh}{\tilde{h}}
\newcommand{\prt}{\partial}

\newcommand{\rgl}{\rangle}
\newcommand{\lgl}{\langle}
\newcommand{\ep}{\epsilon}
\begin{document}

\title{Accelerator dynamics of a fractional kicked rotor}

\author{Alexander Iomin }

\affiliation{Department of Physics, Technion, Haifa, 32000, 
Israel}

\date{\today}
\begin{abstract}
It is shown that the Weyl fractional derivative can quantize 
an open system.
A fractional kicked rotor is studied in the framework of the
fractional Schr\"odinger equation. The system is described by 
the 
non-Hermitian Hamiltonian by virtue of the Weyl fractional 
derivative. Violation of space symmetry leads to acceleration 
of the orbital momentum. Quantum localization saturates this 
acceleration, such 
that the average value of the orbital momentum can be a 
direct current and the system behaves like a ratchet. 
The classical counterpart is a nonlinear kicked rotor with 
absorbing boundary conditions. 

\end{abstract}

\pacs{05.45.Mt, 05.40.-a}

\maketitle

Application of fractional calculus to quantum 
processes is a new approach to the study of fractional properties of 
quantum 
phenomena \cite{kusnezov,laskin1,WBG,laskin2,hermann,levy}.
In this Report we consider a quantum chaotic dynamics of  
a fractional kicked rotor (FKR). The Hamiltonian of the system is
\be\label{H_fkr}
\hat{H}=\hat{\clT}+\ep\cos x\sum_{n=-\infty}^{\infty}
\delta(t-n)\, ,
\ee
where $\ep$ is an amplitude of 
the periodic perturbation which is a train of $\delta$ kicks. 
The kinetic part of 
the Hamiltonian is modeled by 
the fractional Weyl derivative 
\be\label{h0}
\hat{\clT}=(-i\hh)^{\alpha}\clW^{\alpha}/\alpha\, ,
\ee 
where $\hh$ is a dimensionless Planck constant, and 
$\alpha=2-\beta$ with $ 0<\beta<1$. When $\alpha=2$ 
Eq.~(\ref{H_fkr}) is the 
quantum kicked rotor \cite{cas79}. For a periodic function 
$f(x)=\sum \bar{f}_ke^{-ikx}$, the Fourier transform property 
determines 
the fractional Weyl derivative $\clW^{\alpha}$ in the 
following simplest 
way (see \cite{WBG}, ch. 4.3)
\be\label{WD} 
\clW^{\alpha}f(x)=\sum_{n=
-\infty}^{\infty}(-ik)^{\alpha}\bar{f}_ke^{-ikx}\, .
\ee
Since only periodic functions are considered here, this 
oversimplified definition is sufficient without the burden 
of fractional calculus details \cite{add1}. Thus, the 
kinetic term in the Hamiltonian (\ref{H_fkr}) is defined on 
the basis $|k\rgl=e^{ikx}/\sqrt{2\pi}$ as follows:
\be\label{clT} 
\hat{\clT}|k\rgl=\clT(k)|k\rgl=
\frac{(\hh k)^{2-\beta}}{2-\beta}|k\rgl\, .
\ee
This non-Hermitian operator  has complex eigenvalues for 
$k<0$, which are defined on the complex plain with a cut from 
$0$ to $-\infty$, such that $1^{-\beta}=1 $ and 
$(-1)^{-\beta}=\cos\beta\pi-i\sin\beta\pi $, and 
therefore, 
$k^{-\beta}=|k|^{-\beta}e^{-i\pi\beta(k)}$, where
$\beta(k)=\beta[1-{\rm sgn}(k)]/2$ ~\footnote{When 
$\alpha>2$, one chooses $(-1)=e^{-i\pi}$, such that 
$(-1)^{\beta}=e^{-i\pi\beta}$.}. It is worth mentioning that 
the fractional derivative in Eq. (\ref{h0}) appears naturally 
in quantum lattice dynamics with long range interaction
\cite{laskin2},
where $(-ik)^{\alpha}$ is a particular case of polylogarithm 
(see Appendix in Ref. \cite{laskin2}).

A quantum map for the wave function $\psi(x,t)$ is
\be\label{QM}
\psi(x,t+1)=\hat{U}\psi(x,t)\, ,
\ee
where the evolution operator on the period 
\be\label{evol}
\hat{U}=\exp\left[\frac{-i\ep\cos x}{\hh}\right]
\exp\left[\frac{-i\hat{\clT}}{\hh}\right]
\ee
describes free dissipative motion and then a kick. 
Dynamics of the FKR is studied numerically, where Eq. 
(\ref{clT}) enables one to use the fast Fourier transform as 
an efficient way to iterate the quantum map (\ref{QM}).
A specific property of this Hamiltonian dynamics is quantum
dissipation resulting in probability leakage and 
described by the survival probability 
\be\label{surv}
P(t)=\lgl\psi(t)|\psi(t)\rgl=\sum_{n=-\infty}^{\infty}|f_n|^2 \, ,
\ee
where $|f_n|^2$ is probability of level occupation at time $t$.
The initial occupation is $f_n(t=0)=\delta_{n,0}$.
Another specific characteristic is the nonzero mean value of 
the orbital momentum 
\be\label{av_p}
\lgl p(t)\rgl=\hh\frac{\sum_nn|f_n(t)|^2}{\sum_n|f_n(t)|^2}\, ,
\ee
due to the asymmetry of the quantum kinetic term $\hat{\clT}$. 
Results of the numerical study of the quantum map (\ref{QM})
are shown in Figs. 1-4. The quantum dissipation leads to the 
asymmetrical distribution of the level occupation $|f_n(t)|$ 
(see Fig. 1) that results in the nonzero first moment of 
the orbital momentum $\lgl p\rgl\sim t^{\gamma_1}$ in Fig.~2.  
Quantum localization saturates the acceleration with time. 
This accelerator dynamics is 
accompanied by the power law decay of the survival probability 
$P(t)\sim t^{-\gamma_2}$ with the exponent $\gamma_2\approx 0.71$
shown in Fig. 3, and then the decay rate increases with the time
due to quantum effects. 
Quantum localization affects
strongly both $\gamma_1$ and $\gamma_2$. By increase of the 
quantum parameter, when $\hh=0.76$, the exponent $\gamma_1 $ 
approaches zero (in Fig. 4 the slope is $10^{-5}$), and the 
survival probability decays at the rate $\gamma_2\approx 0.99$.

\begin{figure}
\begin{center}
\epsfxsize=8cm
\leavevmode
    \epsffile{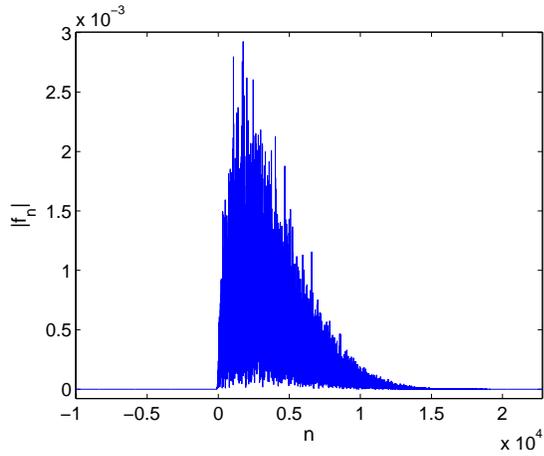}
\caption{Level occupation distribution (after 2000 
iterations) for $\ep=3,~~\beta=0.01, ~~ \hh=0.02$.}
\end{center}
\end{figure}

\begin{figure}
\begin{center}
\epsfxsize=8cm
\leavevmode
    \epsffile{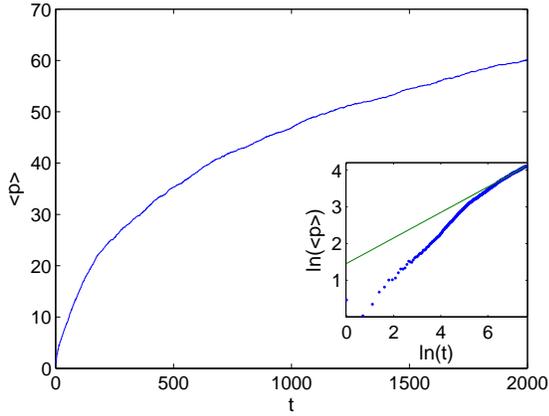}
\caption{Acceleration of the average orbital momentum for the 
same parameters as in Fig.~1. The insert is the log-log plot, 
and the solid line corresponds to $\gamma_1=0.35$ obtained by 
the least squared calculation. } 
\end{center}
\end{figure}

\begin{figure}
\begin{center}
\epsfxsize=8cm
\leavevmode
    \epsffile{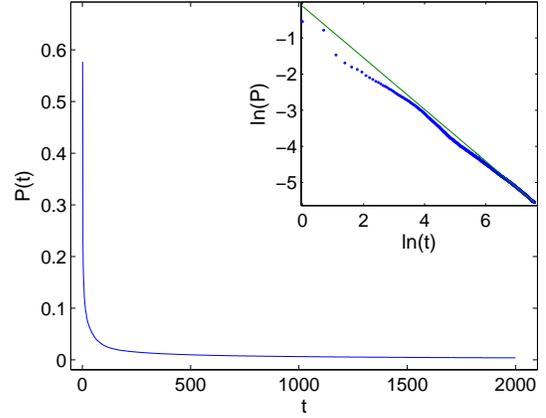}
\caption{Decay of the survival probability  $P(t)$.
The insert is the log-log plot, and the solid 
line corresponds to $\gamma_2=0.71$ obtained by the least 
squared 
calculation. } 
\end{center}
\end{figure}

\begin{figure}
\begin{center}
\epsfxsize=8cm
\leavevmode
    \epsffile{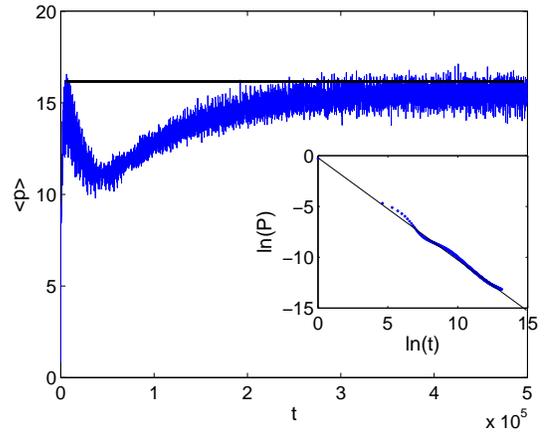}
\caption{Quantum saturation of $\lgl p(t)\rgl $ due to 
localization when $\hh=0.76,~\beta=0.05$ and the same $\ep$ 
as in Figs 1-3.  
The slope of the solid line is $10^{-5}$. The insert shows the 
power low decay of 
$P(t)\sim t^{-\gamma_2}$ with $\gamma_2\approx 0.99$ due to 
the linear interpolation. 
}
\end{center}
\end{figure}

To understand the obtained numerical results and the physical 
relevance of the fractional Schr\"odinger equation, the 
classical limit $\hh\rightarrow 0$ is performed in the Wigner 
representation. Thus, the system is described by the Wigner 
function $W(x,p,t)$ which is a 
$c$-number projection of the density matrix in the Weyl rule of 
association between $c$-numbers and operators.
The Weyl transformation of an arbitrary operator function 
$G(\hat{x},\hat{p})$ is \cite{agarwal,wigner}
\be\label{weyl1}
F(x,p)=\tr\left[G(\hat{x},
\hat{p})\Delta(x-\hat{x},p-\hat{p})\right]\, ,
\ee
where $F(x,p)$ is a $c$-number function and 
$\Delta(x-\hat{x},p-\hat{p})$
is a projection operator which acts as the two dimensional 
Fourier 
transform. For the  cylindrical phase space the projection 
operator is \cite{5_7}
\be\label{weyl2}
\Delta(x-\hat{x},p-\hat{p})=\sum_{m=-\infty}^{\infty}\frac{1}{2\pi}
\int_{-\pi}^{\pi}d\xi e^{im(x-\hat{x})+i\xi(p-\hat{p})}\, .
\ee
This operator determines an inverse transform as well:
\be\label{weyl3}
G(\hat{x},\hat{p})=\sum_{k=-\infty}^{\infty}\frac{1}{2\pi}
\int_{-\pi}^{\pi}F(x,\hh k)\Delta(x-\hat{x},\hh k-\hat{p})\, ,
\ee
where $p=\hh k$. The quantum map for the density matrix 
$\hat{\rho}(t)$ is
\be\label{rhoQM}
\hat{\rho}(t+1)=\hat{U}^{\dag}\hat{\rho}(t)\hat{U}\, .
\ee
Therefore, evolution of the Wigner function 
\[W(t,x,p)=\tr
\left[\hat{\rho}(t)\Delta(x-\hat{x},p-\hat{p})\right]\, ,\]
for the period determined by the map (\ref{rhoQM}), is
\bea\label{wigf}
&W(t+1,x,p)=\tr\left[\hat{U}^{\dag}\hat{\rho}(t)\hat{U}
\Delta(x-\hat{x},p-\hat{p})\right]] \nonumber \\
&=\sum_{k'=-\infty}^{\infty}\int_0^{2\pi}
K_{\hh}(x,p|x',p')W(t,x',p')dx' \, , 
\eea
where $K_{\hh}(x,p|x',p')$ is Green's function for the period
\bea\label{greenf}
&K_{\hh}(x,p|x',p')=\sum_m\frac{1}{2\pi}\int_{-\pi}^{\pi}
e^{im(x-x'+\xi')}e^{i\xi'(p'-p)} \nonumber \\
&\times\exp\left[\frac{i}{\hh}
\clT^*(p+\hh m/2)-\frac{i}{\hh}\clT(p-\hh 
m/2)\right]
\nonumber \\
&\times\exp\left[\frac{i\ep}{\hh}\cos(x'+\hh\xi'/2)
-\frac{i\ep}{\hh}\cos(x'-\hh\xi'/2)\right]\, .
\eea
The trace is $\tr[...]=\sum_k\lgl k|...|k\rgl$. 
In the classical limit $\hh\rightarrow 0$,  we obtain in Eq. 
(\ref{greenf}) that the difference 
of the perturbations in the exponential is $-i\ep\cos x$, 
while the difference of the kinetic terms is
$imp^{1-\beta}\equiv im\omega(p)$ for $p>0$ and 
$-2\sin(\beta\pi)\clT(|p|)/\hh$ for $p<0$. The last term 
diverges at $\hh=0$ and yields  identical zero for Green's 
function $K_{\hh=0}(p<0)\equiv 0$. Thus, the classical Green 
function
\be\label{cl_green}
K_{\hh=0}(x,p|x'p')=\Theta(p)\delta(x-x'-\omega(p))
\delta(p-p'-\ep\sin x')
\ee
corresponds to the classical map ${\cal M}$
\be\label{clmap}
p_{n+1}=p_n+\ep\sin x_n, ~~ x_{n+1}=x_n+\omega(p_{n+1})
\ee 
of the nonlinear kicked rotor with the 
nonlinear frequency $\omega(p)$, and absorbing boundary 
conditions for $p<0$, that the Heaviside function $\Theta(p)$ 
reflects.

Therefore, the fractional Hamiltonian (\ref{H_fkr}) 
corresponds to the open system 
Eqs.~(\ref{cl_green}),(\ref{clmap}).
Chaotic dynamics of this open system takes place in the upper 
half of the cylindrical phase space. The stability property 
is determined by the trace of the linearized map $\prt{\cal M}$ 
\be\label{tracdT}
\tr\left[\prt T\right]=2+\ep(1-\beta)p^{-\beta}\cos x \, .
\ee
For any $\ep$ there are stable regions 
$\{\Delta x,\Delta p\}$ determined by the locus of elliptic  
points $\{x_e,p_e\}$ (see Fig. 5)
\be\label{elliptic}
x_e=\arccos\left(-\frac{2p_e^{\beta}}{\ep(1-\beta)}\right)\, .
\ee
The presence of this infinite regular elliptic island  
structure, which leads to the stickiness of chaotic 
trajectories, also results in the power law decay of the
survival probability for the quantum counterpart in Fig.~3.
This power law phenomenon due to quantum tunneling has been an 
issue of extensive studies in quantum chaos \cite{tunnel}.

Quantum localization leads to the exponential restriction of 
the initial 
profile spreading in the orbital momentum space from above. 
This property results in saturation of acceleration of $\lgl 
p(t)\rgl$; namely, at $t\rightarrow\infty$ it follows that 
$\lgl p(t)\rgl\rightarrow {\rm const}$. Such a behavior is 
found for $\hh=0.76$ and $\beta=0.05$.
In Fig.~4 one sees a direct current of $\lgl p(t)\rgl$
for $5\cdot 10^5$ iterations and $K=3$. This double impact of 
asymmetric absorption and quantum localization leads, 
asymptotically, to a quantum like ratchet
which differs from quantum one obtained on a classical chaotic 
attractor \cite{casati}.

It is worth mentioning that in the class of periodic 
functions, eigenvalues of the unperturbed
Hamiltonian $\clT$ coincide with
$\hat{H}_0(\hat{p})=\left(-i\hh\frac{\prt}{\prt 
x}\right)^{\alpha}$,
and have the same classical limit of Eq. (\ref{cl_green}). 
This local 
derivative has the classical counterpart with the Hamiltonian
$H_0(p)=p^{\alpha}$ which does not coincide with Eq. 
(\ref{cl_green}).
Namely, the Hamiltonian $H_0(p)$ is the classical 
system with dissipation for $p<0$, while the map ${\cal M}$ in 
Eqs. (\ref{cl_green}) and (\ref{clmap}) is the open system where a 
particle is set apart from the dynamics for $p<0$.

Fractional Schr\"odinger equation 
\be\label{fse}
i\hh\prt_t\psi=(-i\hh)^{\alpha}\clW^{\alpha}\psi
\ee
describes quantum {\it dissipative Hamiltonian} dynamics. The 
classical counterpart is a nonlinear motion with 
dispersion $\omega(p)$ realized on the 
upper half plain of the phase space with absorption in the 
lower half plain. It has well defined physical meaning. 
Therefore, the fractional Schr\"odinger equation (\ref{fse}) 
can be a generalized approach for any functions for which the 
Fourier transform is valid. In this case, the opposite 
classical--to--quantum transition can be 
performed by determining the Heaviside function in Eq. 
(\ref{greenf})
\[e^{i\omega(p)z}\Theta(p)=\lim_{\hh\rightarrow 0} 
\exp\left[\frac{i}{\hh}\clT^*(p+\frac{\hh z}{2})-
\frac{i}{\hh}\clT(p-\frac{\hh z}{2})\right]\, ,\]
where $\clT(p)$ is uniquely defined by the condition 
$\omega(p)=\clT^{\prime}(p)$.
Thus, fractional derivatives quantize    
classical open systems in the framework of the non-Hermitian 
Hamiltonians.
\begin{figure}[t]
\begin{center}
\epsfxsize=8cm
\leavevmode
    \epsffile{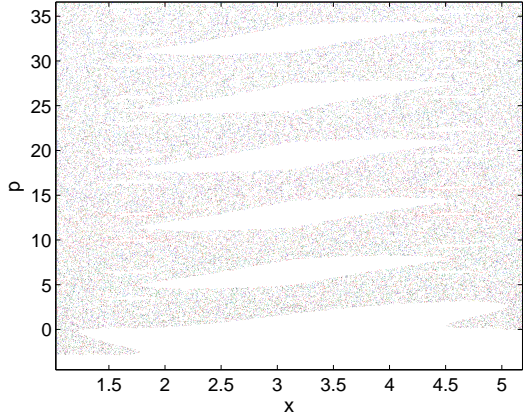}
\caption{Phase portrait of the classical map without 
absorption, after 20000  
iteration of 15 initial conditions for $\ep=3$ and 
$\beta=0.01$.}
\end{center}
\end{figure}

The author would like to thank Prof. G.M. Zaslavsky for 
inspiration, and Prof. R.J. Dorfman and Prof. S. Fishman for 
helpful discussions.
The hospitality of the Max--Planck--Institute of Physics of 
Complex Systems (Dresden), where a part of the work was 
performed, is gratefully acknowledged. 
This research was supported by the Israel Science Foundation.


\begin{thebibliography}{99}

\bibitem{kusnezov} D. Kusnezov, A. Bulgac, and G. D. Dang,
Phys. Rev. Lett. {\bf 82}, 1136 (1999). 

\bibitem{laskin1} N. Laskin, Chaos {\bf 10}, 780 (2000).

\bibitem{WBG} B.J. West, M. Bologna, and P. Grigolini, {\it 
Physics of Fractal Operators}, (Springer, New York, 2002).

\bibitem{laskin2}  N. Laskin and  G. Zaslavsky, 
arXiv:nlin.SI/0512010.

\bibitem{hermann} R. Hermann, arXiv:math-ph/0510099.

\bibitem{levy} see also M. Chaichian A. Demichev A. {\it Path 
integrals in physics: Stochastic process and quantum 
mechanics} Vol. 1, (2001).

\bibitem{cas79} G. Casati, B.V. Chirikov, F.M. Izrailev and 
J. Ford, in {\em Stochastic Behavior in Classical and Quantum 
Hamiltonian Systems}, eds. G. Casati and J. Ford (Springer, 
New York, 1979).

\bibitem{add1} Fractional derivation was developed as a 
generalization of integer order derivatives and is defined as 
the inverse operation to the fractional integral. Fractional 
integration of the order of $\alpha$ is defined by the 
operator (see {\em e.g.}, \cite{WBG} and [I. Podlubny, {\em 
Fractional Differential Equations}, (Academic Press, San 
Diego, 1999)]) \\ 
${}_aI_x^{\alpha}f(x)=
\frac{1}{\Gamma(\alpha)}\int_a^xf(y)(x-y)^{\alpha-1}dy\, 
, $ \\
where $\alpha>0,~x>a$ and  $\Gamma(z)$ is the Gamma function. 
Therefore, the fractional derivative is the inverse operator  
to ${}_aI_x^{\alpha}$ as $ 
{}_aD_x^{\alpha}f(x)={}_aI_x^{-\alpha}$ and
${}_aI_x^{\alpha}={}_aD_x^{-\alpha}$. Its explicit form is \\
${}_aD_x^{-\alpha}=
\frac{1}{\Gamma(-\alpha)}\int_a^xf(y)(x-y)^{-1-\alpha}dy\, 
. $ \\
For arbitrary $\alpha>0$ this integral diverges, and as a 
result of a regularization procedure, there are two 
alternative definitions of ${}_aD_x^{-\alpha}$. For an 
integer $n$ defined as $n-1<\alpha<n$, one obtains 
the Riemann-Liouville fractional derivative of the form \\ 
${}_aD_{RL}^{\alpha}f(x)=(d^n/x^n){}_aI_x^{n-\alpha}f(x)$,\\
and fractional derivative in the Caputo form \\
${}_aD_{C}^{\alpha}f(x)= {}_aI_x^{n-\alpha}f^{(n)}(x)$. \\
There is no constraint on the lower limit $a$. For example, 
when $a=0$, one has  \\
${}_0D_{RL}^{\alpha}x^{\beta}=\frac{x^{\beta-\alpha
\Gamma(\beta+1}}{\Gamma(\beta+1-\alpha)} $\\
and 
$ {}_aD_{C}^{\alpha}f(x)=
{}_0D_{RL}^{\alpha}f(x)-\sum_{k=0}^{n-1}f^{(k)}(0^+)
\frac{x^{k-\alpha}}{\Gamma(k-\alpha+1)}\, ,$
and ${}_aD_{C}^{\alpha}[1]=0$, while 
${}_0D_{RL}^{\alpha}[1]=x^{-\alpha}/\Gamma(1-\alpha)$. \\
When $a=-\infty$, the resulting Weyl derivative is \\ 
$\clW^{\alpha}\equiv{}_{-\infty}D_{W}^{\alpha}=
{}_{-\infty}D_{RL}^{\alpha}=
{}_{-\infty}D_{C}^{\alpha}\, .$ \\
One also has ${}_{-\infty}D_{W}^{\alpha}e^x=e^x$
This property is convenient for the Fourier transform 
$\clF\left[\clW^{\alpha}f(x)\right]=(ik)^{\alpha}\bar{f}(k)$, 
where $\clF[f(x)]=\bar{f}(k)$.

\bibitem{wigner} E.P. Wigner, Phys. Rev. {\bf 40}, 749 (1932).

\bibitem{agarwal} G.S. Agarwal and E. Wolf, Phys. Rev. D {\bf 
2}, 2161 (1970).

\bibitem{5_7} G.P. Berman and A.R. Kolovsky, Physica D {\bf 
17}, (1985);
G.P. Berman, F.M. Izrailev, and A.R. Kolovsky, Physica A {\bf 
152}, 237 (1988); G.P. Berman, A.R. Kolovsky, F.M. Izrailev, 
and A.M. Iomin, Chaos {\bf 1}, 220 (1991).

\bibitem{casati} G.G. Carlo, G. Benenti, G. Casati, and D.L. 
Shepelyansky, Phys. Rev. Lett. {\bf 94}, 16410 (2005).

\bibitem{tunnel} G. Casati, G. Maspero, and D.L. Shepelyansky,
Phys. Rev. Lett. {\bf 82}, 524 (1999);
R. Ketzmerick, L. Hufnagel, F. Steinbach, and M. Weiss,
Phys. Rev. Lett. {\bf 85}, 1214 (2000); A. Iomin, S. Fishman, 
and G.M. Zaslavsky, Phys. Rev. E. {\bf 65}, 036215 (2002).



\end{thebibliography}
\end{document}